\documentclass[twocolumn,english]{revtex4}
\usepackage[T1]{fontenc}
\usepackage[latin1]{inputenc}
\usepackage{graphicx}
\usepackage{amssymb}

\makeatletter

\date{\today}

\usepackage{babel}
\makeatother
\begin{document}

\title{Fluctuation induced first-order phase transitions in a dipolar Ising
ferromagnet slab}

\author{Rafael M. Fernandes$^{1,2}$ and Harry Westfahl Jr.$^{1}$}

\address{\emph{1) Laboratório Nacional de Luz Síncrotron, Caixa Postal 6192,
13084-971, Campinas, SP, Brazil}}

\address{\emph{2) Instituto de Física {}``Gleb Wataghin'', Universidade
Estadual de Campinas, 13083-970, Campinas, SP, Brazil}}

\begin{abstract}
We investigate the competition between the dipolar and the exchange
interaction in a ferromagnetic slab with finite thickness and finite
width. From an analytical approximate expression for the Ginzburg-Landau
effective Hamiltonian, it is shown that, within a self-consistent
Hartree approach, a stable modulated configuration arises. We study
the transition between the disordered phase and two kinds of modulated
configurations, namely, striped and bubble phases. Such transitions
are of the first-order kind and the striped phase is shown to have
lower energy and a higher spinodal limit than the bubble one. It is
also observed that striped configurations corresponding to different
modulation directions have different energies. The most stable are
the ones in which the modulation vanishes along the unlimited direction,
which is a prime effect of the slab's geometry together with the competition
between the two distinct types of interaction. An application of this
model to the domain structure of MnAs thin films grown over GaAs substrates
is discussed and general qualitative properties are outlined and predicted,
like the number of domains and the mean value of the modulation as
functions of temperature.
\end{abstract}
\maketitle

\section{Introduction}

Magnetic phase transitions in materials with finite spacial dimensions
is still a subject with many aspects to be understood and to be investigated
theoretically. In these systems, one finds the competition between
a strong, short-range interaction (exchange) and another weak, long-range
one (dipolar), from which a modulated stable configuration is expected
to outcome \cite{science}. Due to the variety of possible modulated
patterns, like striped, bubble and intermediate shapes, many different
complex domain structures are likely to be seen. This phenomenon is
observed not only in magnetic materials \cite{doniach,allen}, but
also in other systems characterized by the same kind of competition
between an organizing local interaction and a frustrating long-range
interaction \cite{science}: for example, spontaneous modulation of
mesoscopic phases is found in biological systems, amphifilic solutions
\cite{wu} , Langmuir monolayers \cite{langmuir} and block copolymers
\cite{fredrickson}.

There are, indeed, several analytical and numerical studies in the
literature investigating size effects on the critical behaviour of
magnetic systems \cite{finito_pbc1,finito_pbc2,finito_pbc3}. However,
the majority of them deals with periodic boundary conditions, and
not free boundaries, which comes to be the case for many materials.
Even the works related to this latter kind of systems \cite{finito_Dbc1,finito_Dbc2,finito_Dbc3}
do not take into account the weak, long-range (dipolar) interaction,
which changes drastically the underlying physics. As shown by Garel
and Doniach \cite{doniach}, the inclusion of the dipolar interaction
in the Ginzburg-Landau effective Hamiltonian of a magnetic slab with
infinite width leads to a minimum in the Fourier space characterized
by a non-zero wave vector. This is responsible not only for instability
towards the spacial homogeneous phase but also for the existence of
a large volume, in the Fourier space, for fluctuations of the order
parameter to take place and induce a first-order transition (Brazovskii
transition \cite{braso}). Therefore, to achieve a more complete understanding
of finite magnetic systems, it is necessary not only to consider the
finiteness of the them but also the fluctuations of the order parameter. 

To explain some properties of many real materials, size effects are
in fact necessary. For example, concerning non-magnetic systems, Huhn
and Dohm have proposed that size effects are responsible for the temperature
shift of the specific heat maximum in confined $\mathrm{He^{4}}$
\cite{finito_Dbc2}. In what concerns magnetic systems, a material
that has been deeply experimentally investigated in recent years and
in which size effects may play a significant role is MnAs thin films
grown on GaAs \cite{mnas_prl}. The reason why this heterostructure
is calling so much attention is due not only to its academical appeal
but also to its possible application as a spintronic device \cite{tanaka}.
In contrast with bulk MnAs, which presents an abrupt transition from
the low temperature hexagonal (ferromagnetic) $\alpha$ phase to the
high temperature orthorhombic (paramagnetic) $\beta$ phase \cite{bean},
the MnAs:GaAs films show a wide region of coexistence between $\alpha$
and $\beta$ phases from approximately $0\,^{\circ}\mathrm{C}$ to
$50\,^{\circ}\mathrm{C}$ \cite{kaganer,ney,engel,paniago,iikawa}.
In this region, periodic stripes subdivided in ferromagnetic and paramagnetic
terraces arise. X-ray diffraction experiments \cite{paniago} and
microscopy measurements \cite{engel} have shown that, while the temperature
varies, the width of the ferro and paramagnetic terraces change, but
the stripes remain with the same periodic width. These experiments
have also brought out the terraces morphology, showing the complex
phases inside the ferromagnetic terraces. As their width is of the
same order of magnitude as their thickness, both spacial limitations
are important to understand their internal domain structure.

Here is an outline of the article: in Section 2, we construct an expression
for the Ginzburg-Landau effective Hamiltonian of a dipolar Ising ferromagnetic
slab with finite width and finite thickness, considering Dirichlet
boundary conditions (i.e., vanishing of the order parameter at the
walls). We show that a modulated phase arises as the ordered one,
and is represented by a dotted semi-ellipsis in the Fourier space
as a result of frustration and Dirichlet boundary conditions. In Section
3, we apply a self-consistent Hartree calculation to take into account
the fluctuations of the order parameter around the region of minimum
energy. Generalizing the original method developed by Brazovskii \cite{braso}
to the case of this finite system, we calculate the free energy profiles
for two different types of modulation: striped and bubble phases.
We show that the striped phases are more stable than the bubble ones
and also that the energy degeneracy concerning the region of minimum
energy is broken along the finite direction. In Section 4, we discuss
the application of the model to the real case of MnAs:GaAs films.
Although in such systems the magnetization is rather vectorial than
Ising-type, general qualitative properties due to the nature of the
interactions and to the slab's geometry can be obtained. Section 5
is devoted to the final remarks and followed by an appendix where
details of some calculations are explicitly derived.

\section{Ginzburg-Landau for the finite slab}

We consider a slab with thickness $D$ ($z$ axis), width $d$ ($x$
axis) and no limitation along the $y$ axis; the magnetization is
supposed to point only to the $\hat{z}$ direction (Ising model) and
to depend upon $x$ and $y$ only (uniform along $\hat{z}$). This
is, indeed, a very simplified model of a ferromagnetic stripe on the
MnAs:GaAs coexistence region, but we will postpone this discussion
until the last section. Our purpose in this section is to obtain a
two-dimensional Ginzburg-Landau to describe the system; first, consider
the well-known mean field expansion of the free energy due to the
exchange interaction \cite{negele}:

\begin{equation}
F_{exch}[m]=D\int d^{2}rf_{exch}\,,\label{acao_troca}\end{equation}
where

\begin{equation}
f_{exch}=\frac{T_{c}}{16a}\left|\vec{\nabla}m\right|^{2}+\frac{(T-T_{c})}{2a^{3}}m^{2}+\frac{T_{c}}{12a^{3}}m^{4}\,.\label{energia_troca}\end{equation}

$T_{c}$ is the Ising ferromagnetic critical temperature, $a$ is
the lattice parameter and $m(\vec{r})$, the scalar order parameter,
is the coarse-grained spin in the position $\vec{r}=(x,y)$. We are
denoting, through all this article, the integrals over the region
limited by the plane of the slab as:

\[
\int d^{2}r=\int_{-\infty}^{\infty}\int_{0}^{d}dxdy\,.\]

For the sake of simplicity, we consider a cubic lattice. In the long
wavelength limit, the actual crystallographic structure will not change
significantly the basic physical properties of the system. In this
limit, we can calculate the dipolar contribution to the total energy
$f_{dip}$ using just Maxwell equations.

To obtain $f_{dip}$, we express the magnetization $M(\vec{r})$ at
the position $\vec{r}$ in terms of its Fourier components as

\begin{equation}
M(\vec{r})=\frac{g\mu_{B}}{a^{3}}\sum_{n>0,q_{y}}m_{n,q_{y}}\sin\left(\frac{n\pi x}{d}\right)e^{-iq_{y}y}\,,\label{magnet_fourier}\end{equation}
where $g$ is the gyromagnetic factor, $\mu_{B}$ is the Bohr magneton
and:

\begin{equation}
m_{n,q_{y}}=\frac{2}{dL_{y}}\int_{0}^{d}\int_{-\infty}^{\infty}m(\vec{r})\sin\left(\frac{n\pi x}{d}\right)e^{iq_{y}y}dxdy\,.\label{coef_fourier}\end{equation}

In expression (\ref{magnet_fourier}), the sine term appears as a
consequence of the boundary condition that the magnetization vanishes
at the edges of the slab (Dirichlet boundary conditions). In what
concerns MnAs:GaAs thin films, this condition approximates the fact
that, in the coexistence region, the ferromagnetic terraces are succeeded
by paramagnetic ones. 

As we show in Appendix A, the magnetostatic energy of the arbitrary
configuration (\ref{magnet_fourier}) can be straightforward calculated
from Maxwell equations, yielding:

\begin{widetext}

\begin{equation}
f_{dip}=\left(\frac{g\mu_{B}}{a^{3}}\right)^{2}\sum_{q_{y},n,n'}4\pi^{2}nn'm_{n,q_{y}}m_{n',-q_{y}}p\int_{0}^{\infty}du\frac{\left(1-e^{-\frac{1}{p}\sqrt{u^{2}+(q_{y}d)^{2}}}\right)\left[1+(-1)^{nn'+1}\cos(u)\right]}{\sqrt{u^{2}+(q_{y}d)^{2}}\left(u^{2}-n^{2}\pi^{2}\right)\left(u^{2}-n'^{2}\pi^{2}\right)}\,,\label{desmagnet_energ}\end{equation}

\end{widetext} where the sums are over $n$ and $n'$ with same parity
and where we have denoted the slab's aspect ratio as $p=d/D$. As
we wish a simple model to describe the main physical properties of
the slab, we look for an analytical approximation for (\ref{desmagnet_energ}).
Disregarding the cross terms, which are usually negligible, we have
that the sum of the direct terms is:

\begin{equation}
f_{dip}=\left(\frac{g\mu_{B}}{a^{3}}\right)^{2}\sum_{n,q_{y}}\pi\frac{\left(1-e^{-qD}\right)}{qD}m_{n,q_{y}}m_{n,-q_{y}}\,,\label{energ_analit}\end{equation}
where we denoted the wave vector modulus by:

\begin{equation}
q=\sqrt{\frac{n^{2}\pi^{2}}{d^{2}}+q_{y}^{2}}\,.\label{aux_momento-total}\end{equation}

The accuracy of the approximation (\ref{energ_analit}) depends on
the values of the parameters involved and on the pair $(n,q_{y})$
considered. In the experimental case of interest, namely, the MnAs:GaAs
films in the neighbourhood of the phase transition between the ordered
and disordered phases, this approximation implies in errors less than
$20\%$ as long as $p>0.5$. 

The approximate analytical expression obtained is very similar to
the expression deduced by Garel and Doniach for a slab with infinite
width \cite{doniach}. The only difference is that, in the present
case, the wave vector component along the $x$ direction is discrete
due to the slab's finite width. It is clear that in the limit $d\rightarrow\infty$
we recover the same expression.

Using equation (\ref{energ_analit}), we obtain the following expression
for the total free energy density

\begin{widetext}

\begin{equation}
f_{tot}=\sum_{n>0,q_{y}}\left[\frac{T-T_{c}}{4a^{3}}+f(q)\right]m_{n,q_{y}}m_{n,-q_{y}}+\frac{T_{c}}{96a^{3}}\sum_{\{ n_{i}\},\{ q_{i}\}}m_{\{ n_{1},n_{2},n_{3}\},-q_{1}-q_{2}-q_{3}}m_{n_{1},q_{1}}m_{n_{2},q_{2}}m_{n_{3},q_{3}}\,,\label{energ_total}\end{equation}
 where

\begin{equation}
f(q)=\frac{T_{c}q^{2}}{32a}+\left(\frac{g\mu_{B}}{a^{3}}\right)^{2}\pi\frac{\left(1-e^{-qD}\right)}{qD}\,,\label{energ_momento}\end{equation}
and

\begin{eqnarray}
\left\langle m\right\rangle _{\{ n_{1},n_{2},n_{3}\},-q_{1}-q_{2}-q_{3}} & = & \left(\left\langle m\right\rangle _{n_{1}+n_{2}-n_{3}}+\left\langle m\right\rangle _{n_{1}-n_{2}+n_{3}}+\left\langle m\right\rangle _{-n_{1}+n_{2}+n_{3}}-\left\langle m\right\rangle _{n_{1}+n_{2}+n_{3}}\right.\nonumber \\
 &  & \left.-\left\langle m\right\rangle _{n_{1}-n_{2}-n_{3}}-\left\langle m\right\rangle _{-n_{1}+n_{2}-n_{3}}-\left\langle m\right\rangle _{-n_{1}-n_{2}+n_{3}}\right)_{,-q_{1}-q_{2}-q_{3}}\,.\label{def_quartico}\end{eqnarray}

\end{widetext}

\begin{figure}
\includegraphics[%
  width=0.48\columnwidth]{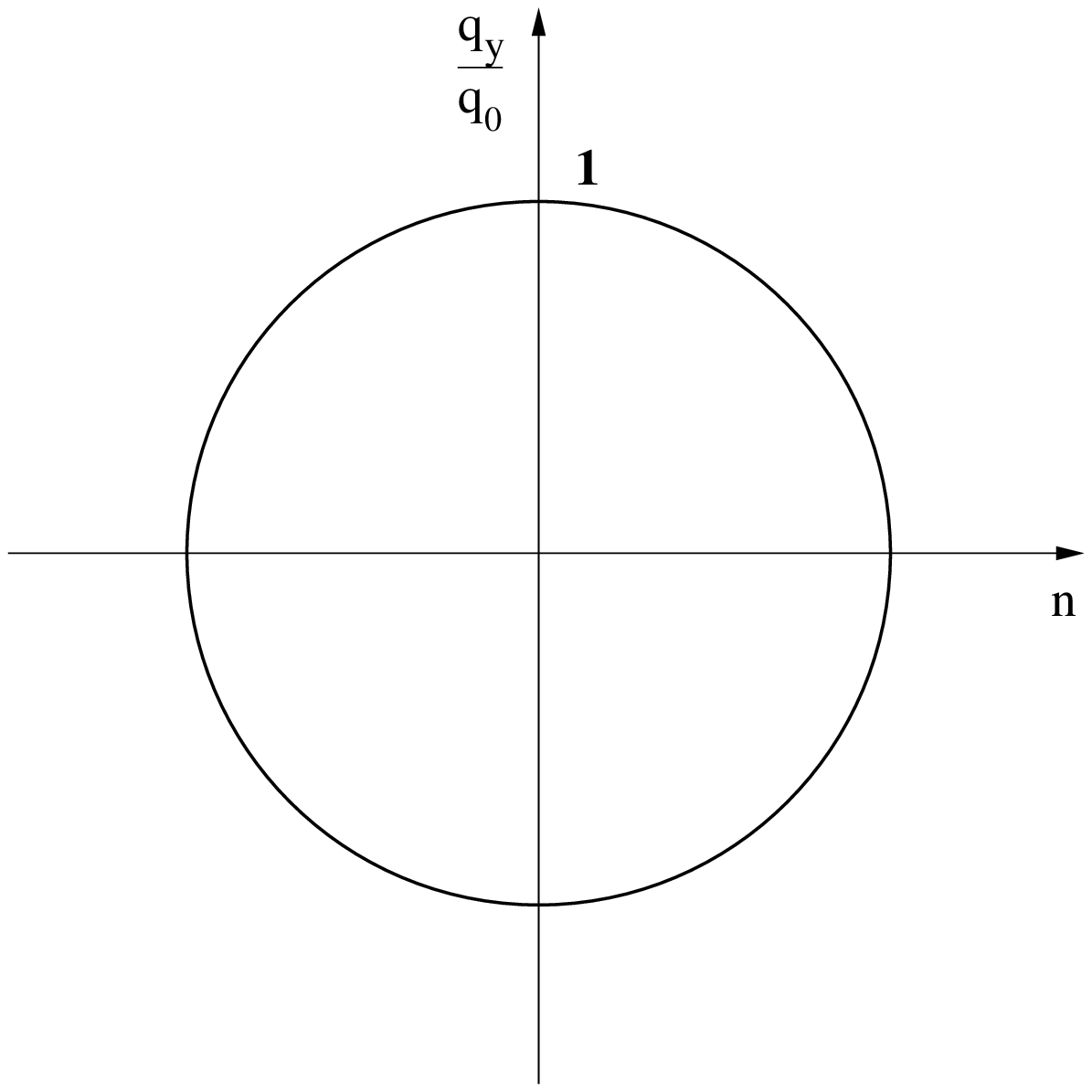}\hfill{}\includegraphics[%
  width=0.48\columnwidth]{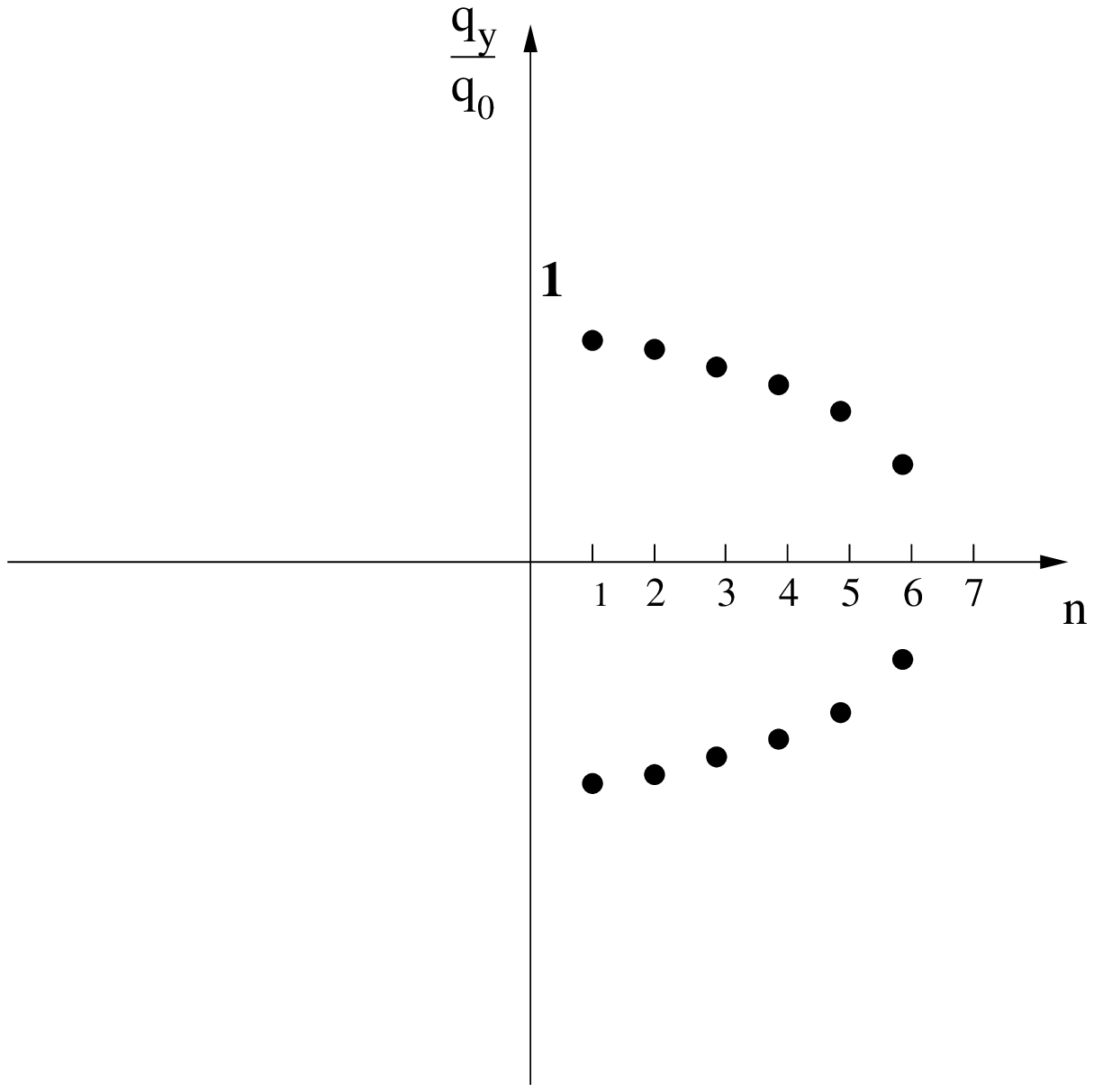}

~(a)~~~~~~~~~~~~~~~~~~~~~~~~~~~~~~~~~~~~~~~(b)

\caption{Region on the momentum (Fourier) space corresponding to the minimum
of the Ginzburg-Landau for the case of a slab with (a) infinite width
and (b) finite width.\label{fig:phase_space}}
\end{figure}

The quartic term is the same as those obtained in previous works about
finite systems with Dirichlet boundary conditions (see, for instance,
\cite{finito_Dbc1,finito_Dbc3}). The effects of the two interactions
mentioned before are evident from (\ref{energ_momento}): while the
$q^{2}$ term, generated by the exchange energy, favours $q=0$ (non-modulated)
configurations, the last term, generated by the dipolar energy, favours
$q\rightarrow\infty$ configurations. The total energy reaches its
minimum value when the wave vector modulus is given by (as long as
$q_{0}D\gg1$):

\begin{equation}
q_{0}=\frac{1}{a}\left(\frac{16\pi g^{2}\mu_{B}^{2}}{T_{c}Da^{2}}\right)^{1/3}\,,\label{minimo}\end{equation}
which means that the most relevant thermodynamic states are characterized
by a non-zero modulation ($q_{0}\neq0$). Expression (\ref{minimo})
is the same as the one obtained by Garel and Doniach in \cite{doniach}
for the minimum energy of the infinite slab; however, in their case,
the phase space of low energy excitations is described a circle in
the Fourier space, whereas in the present situation it is described
by a dotted semi-ellipsis in the $(n,q_{y})$ space, as it is shown
in figure \ref{fig:phase_space}. 

Expanding (\ref{energ_total}) around its minimum, it is straightforward
to obtain the respective partition function:

\begin{equation}
Z=\int\mathcal{D}m\exp\left(-\mathcal{H}[m]\right)\,,\label{particao}\end{equation}
where the Ginzburg-Landau effective Hamiltonian $\mathcal{H}$ is
given by:

\begin{eqnarray}
\mathcal{H}[m] & = & \frac{1}{2}\int\int d^{2}rd^{2}r'm(\vec{r})G_{0}^{-1}(\vec{r},\vec{r}')m(\vec{r}')\nonumber \\
 &  & +\frac{u}{4}\int d^{2}rm^{4}(\vec{r})\,,\label{Ginzb-Land}\end{eqnarray}
and the correlation function $G_{0}$ is written in terms of its Fourier
series as:

\begin{widetext}

\begin{eqnarray}
G_{0}^{-1}(\vec{r},\vec{r}') & = & \frac{1}{\pi d}\sum_{n>0}\int dq_{y}\left.G_{n,q_{y}}^{(0)}\right.^{-1}\sin\left(\frac{n\pi x}{d}\right)\sin\left(\frac{n\pi x'}{d}\right)e^{-iq_{y}(y-y')}\,,\nonumber \\
\left.G_{n,q_{y}}^{(0)}\right.^{-1} & = & \frac{1}{r_{0}+c\left(q-q_{0}\right)^{2}}\,.\label{correlac_funct}\end{eqnarray}

\end{widetext}

The parameters $u$, $r_{0}$ and $c$ appearing in the above expressions
can be written in terms of the microscopic parameters of the systems
as:

\begin{eqnarray}
u & = & \frac{DT_{c}}{3a^{3}T}\,,\nonumber \\
r_{0} & = & \frac{\left(T-T^{*}\right)D}{a^{3}T}\,,\nonumber \\
c & = & \frac{3DT_{c}}{8aT}\,,\label{parameters_GL}\end{eqnarray}
where we defined the shifted critical temperature:

\begin{equation}
T^{*}=T_{c}\left[1-6\left(\frac{\pi g^{4}\mu_{B}^{4}}{16D^{2}a^{4}T_{c}^{2}}\right)^{1/3}\right]\,.\label{shifted_temp}\end{equation}

Therefore, considering the expression for the correlation function,
(\ref{correlac_funct}), it is expected some similarity between this
system and the Brazovskii's model \cite{braso}. The question is if
the reduction in the momentum (Fourier) space, due to the discreteness
of the $x$ components ($q_{x}=n\pi/d$), is able to substantially
change the picture, since Garel and Doniach have shown that, for the
case of an infinite slab, in which $q_{x}$ is continuous, a fluctuation
induced first-order phase transition does occur between the ordered
(modulated) and the disordered phase \cite{doniach}.

\section{Hartree calculation and Brazovskii's procedure}

As we detailed in the previous section, there is a degenerate region
at the Fourier space, expressed by the non-zero wave vector modulus
$q_{0}$, corresponding to the minimum of the Ginzburg-Landau effective
Hamiltonian. Hence, there is a large space for fluctuations of the
order parameter to take place, and a mean-field approach to calculate
the partition function (\ref{particao}) - and its corresponding thermodynamical
properties - is not satisfactory. To deal with that, we generalize
the procedure adopted by Brazovskii \cite{braso} to our finite system.
Such procedure is based on the Hartree self-consistent method, that
consists on replacing the quartic term by an effective quadratic one
\cite{lubenski}:

\begin{equation}
\frac{u}{4}\int d^{2}rm^{4}(\vec{r})\rightarrow\frac{3u}{2}\int d^{2}r\left\langle m^{2}(\vec{r})\right\rangle m^{2}(\vec{r})\,.\label{aprox_hartree}\end{equation}

Then, using the identity

\[
\left\langle m^{2}(\vec{r})\right\rangle =G(\vec{r},\vec{r})+\left\langle m(\vec{r})\right\rangle ^{2}\,,\]
and substituting in (\ref{Ginzb-Land}), it is clear that a self-consistent
equation is obtained for the correlation function:

\begin{eqnarray}
G^{-1}(\vec{r},\vec{r}') & = & G_{0}^{-1}(\vec{r},\vec{r}')+3uG(\vec{r},\vec{r})\delta(\vec{r}-\vec{r}')+\nonumber \\
 &  & +3u\left\langle m(\vec{r})\right\rangle ^{2}\delta(\vec{r}-\vec{r}')\,,\label{auto-consist_real}\end{eqnarray}

where the Dirac delta function is to be understood as belonging to
the intervals $[0,d]$ ($x$ axis) and $[-\infty,\infty]$ ($y$ axis),
and not to $[-\infty,\infty]$ and $[-\infty,\infty]$, as it is usually
assumed.

Following Brazovskii, the dominant contributions for the correlation
function $G(\vec{r},\vec{r}')$ come from the diagonal Fourier components
$G_{(n,q_{y}),(n,-q_{y})}$ such that $\sqrt{n^{2}\pi^{2}/d^{2}+q_{y}^{2}}=q_{0}$.
Therefore, it is useful to write the self-consistent equation on the
Fourier space

\begin{widetext}

\begin{eqnarray}
G_{n,q_{y}}^{-1} & = & \left.G_{n,q_{y}}^{(0)}\right.^{-1}+\frac{3u}{2}\left\langle G\right\rangle +\frac{3u}{4}\left\langle G\right\rangle _{n}+\frac{9u}{4}\sum_{p_{y}}\left\langle m\right\rangle _{n,p_{y}}\left\langle m\right\rangle _{n,-p_{y}}-\frac{3u}{4}\sum_{p_{y}}\left\langle m\right\rangle _{n,p_{y}}\left\langle m\right\rangle _{3n,-p_{y}}+\nonumber \\
 &  & \frac{3u}{4}\sum_{m\neq n,p_{y}}\left\langle m\right\rangle _{m,p_{y}}\left[2\left\langle m\right\rangle _{m,-p_{y}}-\left\langle m\right\rangle _{m+2n,-p_{y}}+\left\langle m\right\rangle _{2n-m,-p_{y}}-\left\langle m\right\rangle _{m-2n,-p_{y}}\right]\,,\label{aux_hartree_geral}\end{eqnarray}

\end{widetext}where

\begin{eqnarray}
\left\langle G\right\rangle _{m} & = & \frac{1}{\pi d}\int dp_{y}G_{m,p_{y}}\,,\nonumber \\
\left\langle G\right\rangle  & = & \sum_{m}\left\langle G\right\rangle _{m}\,.\label{aux_mean_values}\end{eqnarray}

The summation in (\ref{aux_hartree_geral}) is over the region comprehended
by the dotted semi-ellipsis shell whose thickness $\Lambda$ is such
that $\Lambda\ll q_{0}$. In this region, the diagonal Fourier components
$G_{n,q_{y}}$ can be expanded as:

\begin{equation}
G_{n,q_{y}}=\frac{1}{r+c\left(q-q_{0}\right)^{2}}\label{correlac_renorm}\end{equation}
and their mean values can be evaluated to yield

\begin{eqnarray}
\left\langle G\right\rangle _{n} & = & \frac{4q_{0}}{\pi\sqrt{c}\sqrt{r}}\frac{1}{\sqrt{\frac{q_{0}^{2}d^{2}}{\pi^{2}}-n^{2}}}\,,\nonumber \\
\left\langle G\right\rangle  & = & \frac{4q_{0}}{\pi\sqrt{c}\sqrt{r}}\sum_{m=1}^{N\left(q_{0}d/\pi\right)}\frac{1}{\sqrt{\frac{q_{0}^{2}d^{2}}{\pi^{2}}-m^{2}}}\,.\label{mean_values}\end{eqnarray}

Here $N(x)$ denotes the integer closest to $x$ and smaller than
$x$. Therefore, equation (\ref{aux_hartree_geral}) can be written
as:

\begin{widetext}

\begin{eqnarray}
r & = & r_{0}+\frac{\Gamma_{n}u}{\sqrt{r}}+\frac{9u}{4}\sum_{p_{y}}\left\langle m\right\rangle _{n,p_{y}}\left\langle m\right\rangle _{n,-p_{y}}-\frac{3u}{4}\sum_{p_{y}}\left\langle m\right\rangle _{n,p_{y}}\left\langle m\right\rangle _{3n,-p_{y}}+\nonumber \\
 &  & \frac{3u}{4}\sum_{m\neq n,p_{y}}\left\langle m\right\rangle _{m,p_{y}}\left[2\left\langle m\right\rangle _{m,-p_{y}}-\left\langle m\right\rangle _{m+2n,-p_{y}}+\left\langle m\right\rangle _{2n-m,-p_{y}}-\left\langle m\right\rangle _{m-2n,-p_{y}}\right]\,,\label{hartree_geral}\end{eqnarray}

\end{widetext}where:

\begin{equation}
\Gamma_{n}=\frac{6q_{0}}{\pi\sqrt{c}}\sum_{m=1}^{N\left(q_{0}d/\pi\right)}\frac{\left(1+\frac{\delta_{m,n}}{2}\right)}{\sqrt{\frac{q_{0}^{2}d^{2}}{\pi^{2}}-m^{2}}}\,.\label{def_gamma}\end{equation}

It is clear that the self-consistent equation for the renormalized
parameter $r$ , equation (\ref{hartree_geral}), depends on the system
phase through $\left\langle m(\vec{r})\right\rangle $. However, for
any configuration $\left\langle m(\vec{r})\right\rangle $, analogously
to \cite{braso}, the equation does not allow the solution $r=0$,
what implies that the transition from the disordered to this ordered
phase (characterized by the mean value $\left\langle m(\vec{r})\right\rangle $)
is not second-order. Therefore, a first-order transition is expected,
as well as the raise of metastable states (and the respective spinodal
stability limits).

To further investigate how the system achieves all the possible distinct
modulated states, it is useful to calculate the free energy difference
between these configurations and the disordered one. Hence, we use
the same procedure as Brazovskii and consider that the conjugate field
$h$ grows from zero in the disordered phase to a maximum value and
then goes again to zero in the ordered stable phase, characterized
by an amplitude $A\neq0$ \cite{braso} :

\begin{widetext}

\begin{equation}
\Delta F=F_{ord}-F_{desord}=\int_{0}^{A}\frac{dF}{dA'}dA'=\int_{r}^{r_{A}}\left(\sum_{n,q}\frac{\delta F}{\delta\left\langle m\right\rangle _{n,q}}\frac{d\left\langle m\right\rangle _{n,q}}{dA}\right)\frac{dA}{dr'}dr'\,,\label{dif_energy_definition}\end{equation}
where

\[
\frac{\delta F}{\delta\left\langle m\right\rangle _{n,q}}=h_{n,q}\]
is the conjugate field in the Fourier space. A lengthy but straightforward
calculation gives:

\begin{eqnarray}
h_{n,q} & = & G_{n,q}^{-1}\left\langle m\right\rangle _{n,q}+\frac{u}{4}\sum_{m,m',p,p'}\left\langle m\right\rangle _{m,p}\left\langle m\right\rangle _{m',p'}\left\langle m\right\rangle _{\{ m,m',n\},q-p-p'}-\nonumber \\
 &  & \frac{9u}{4}\left\langle m\right\rangle _{n,q}\sum_{p}\left\langle m\right\rangle _{n,p}\left(\left\langle m\right\rangle _{n,-p}-\frac{1}{3}\left\langle m\right\rangle _{3n,-p}\right)-\nonumber \\
 &  & \frac{3u}{4}\left\langle m\right\rangle _{n,q}\sum_{m\neq n,p}\left\langle m\right\rangle _{m,p}\left[2\left\langle m\right\rangle _{m,-p}-\left\langle m\right\rangle _{m+2n,-p}+\left\langle m\right\rangle _{2n-m,-p}-\left\langle m\right\rangle _{m-2n,-p}\right]-\nonumber \\
 &  & \frac{3u}{4}\left\langle G\right\rangle _{3n}\left\langle m\right\rangle _{3n,q}+\frac{3u}{4}\sum_{m\neq n}\left\langle G\right\rangle _{m}\left(-\left\langle m\right\rangle _{n+2m,q}+\left\langle m\right\rangle _{2m-n,q}-\left\langle m\right\rangle _{n-2m,q}\right)\,.\label{conjugated_field}\end{eqnarray}

\end{widetext}

We follow Garel and Doniach \cite{doniach} and study two different
modulated configurations: the striped and the bubble phases.

\subsection*{Striped phases}

The name stripes may be somehow misleading, as genuine stripes cannot
be formed due to the boundary conditions, unless they lie only along
the $y$ direction. In fact, these phases refer to the simplest modulated
configuration that can arise in the finite system:

\begin{eqnarray}
\left\langle m\right\rangle _{n,q} & = & A\delta_{n,n_{0}}\left(\delta_{q,q_{0y}}+\delta_{q,-q_{0y}}\right)\\
\left\langle m\left(\vec{r}\right)\right\rangle  & = & 2A\sin\left(\frac{n_{0}\pi x}{d}\right)\cos\left(q_{0y}y\right)\,,\label{stripe}\end{eqnarray}
where

\[
\sqrt{\frac{n_{0}^{2}\pi^{2}}{d^{2}}+q_{0y}^{2}}=q_{0}\,.\]

\begin{figure}
\includegraphics[%
  width=0.48\columnwidth]{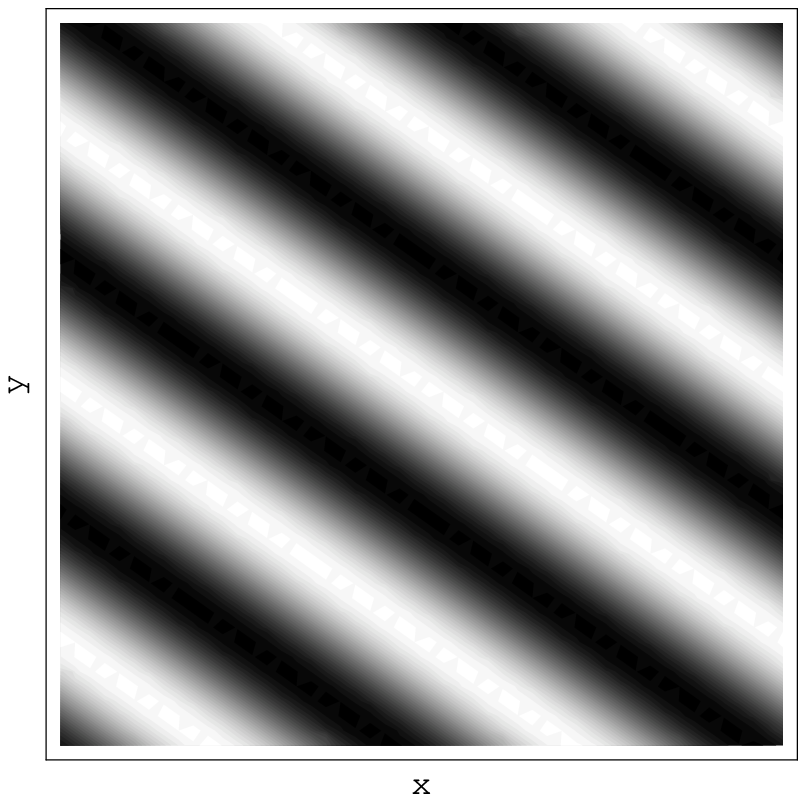}\hfill{}\includegraphics[%
  width=0.48\columnwidth]{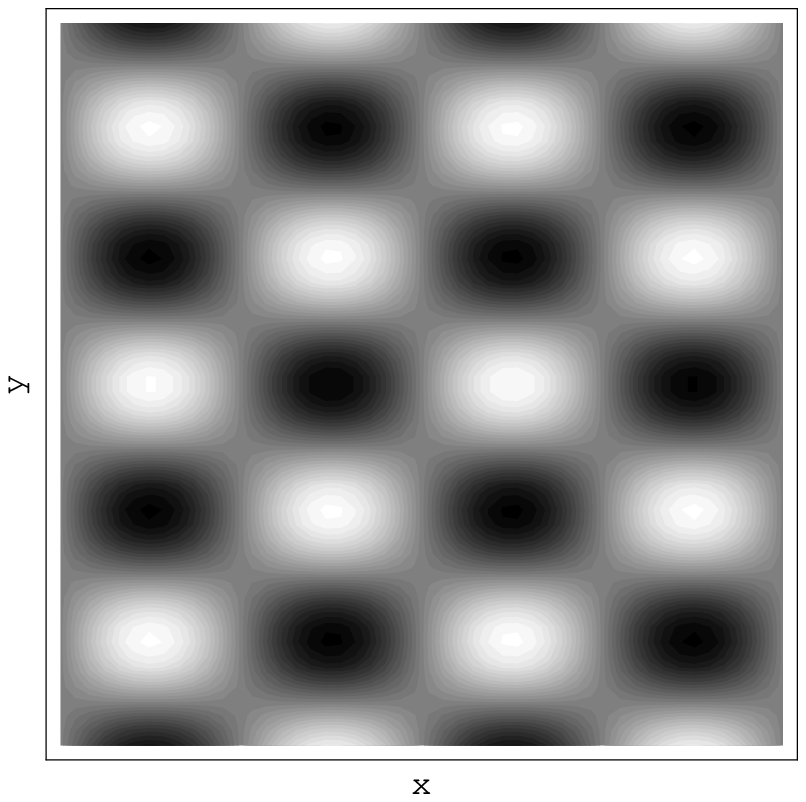}

~(a)~~~~~~~~~~~~~~~~~~~~~~~~~~~~~~~~~~~~~~~(b)

\caption{Contour plot of the order parameter along the slab's plane concerning
the simplest modulated phases that arise in the case of (a) an infinite
slab and (b) a finite slab.\label{fig:stripe}}
\end{figure}

Therefore, the use of the name stripes is to make contact with the
phases of the infinite slab rather than to describe exactly the geometrical
pattern. Figure \ref{fig:stripe} compares the general picture of
the simplest modulated phases of the infinite and of the finite slab. 

Substituting equation (\ref{stripe}) in expressions (\ref{hartree_geral})
and (\ref{conjugated_field}), we obtain the following expressions
for the self-consistent and the state equations:

\begin{eqnarray}
r & = & r_{0}+\frac{\Gamma_{n_{0}}u}{\sqrt{r}}+\frac{9}{2}uA^{2}\,,\nonumber \\
h & = & rA-\frac{9}{4}uA^{3}\,.\label{stripe_eq_aux}\end{eqnarray}

Imposing that the conjugate field vanishes in the ordered phase (denoted
by $r_{A}$), we get:

\begin{equation}
-r_{A}=r_{0}+\frac{\Gamma_{n_{0}}u}{\sqrt{r}}\,.\label{stripe_hartree}\end{equation}

This equation is the same as the one obtained for the case of an infinite
slab, and was considered by Brazovskii in his original work. It implies
that these striped phases can arise as metastable states below the
spinodal stability limit:

\[
r_{spinodal}\approx-1.89\left(u\Gamma_{n_{0}}\right)^{2/3}\,.\]

To obtain the free energy difference between these modulations and
the disordered phase, we take $\left\langle m\left(\vec{r}\right)\right\rangle =0$
in (\ref{hartree_geral}); we get:

\begin{equation}
r=r_{0}+\frac{\Gamma_{n_{0}}u}{\sqrt{r}}\,.\label{disord_hartree}\end{equation}

Using (\ref{dif_energy_definition}), the free energy difference is

\begin{equation}
\Delta F_{s}=\frac{2\left(u\Gamma_{n_{0}}^{4}\right)^{1/3}}{9}\left[-\frac{\rho_{A}^{2}}{2}-\frac{\rho^{2}}{2}-\sqrt{\rho}+\sqrt{\rho_{A}}\right]\,,\label{stripe_energy}\end{equation}
where we defined the auxiliary variables

\begin{equation}
\rho_{i}=\frac{r_{i}}{\left(u\Gamma_{n_{0}}\right)^{2/3}}\,.\label{aux_variable}\end{equation}

The new feature that appears as a consequence of the finiteness of
the system is the dependence of the factor $\Gamma_{n_{0}}$ with
respect to the slab's width $d$ and to the modulation label $n_{0}$,
which indicates what point of the semi-ellipsis is taken to modulate
the system. It is clear that a modulation labeled by $n_{0}$ can
arise only if:

\begin{equation}
\frac{q_{0}d}{\pi}\geq n_{0}\label{condition_n0}\end{equation}
otherwise the semi-ellipsis does not comprehend this specific point.
Such label can be interpreted as the number of {}``spread'' domains
along the $x$ direction, since there are no sharp walls and the magnetization
changes sign continuously from one domain to another. 

\begin{figure}
\includegraphics[%
  width=0.48\columnwidth]{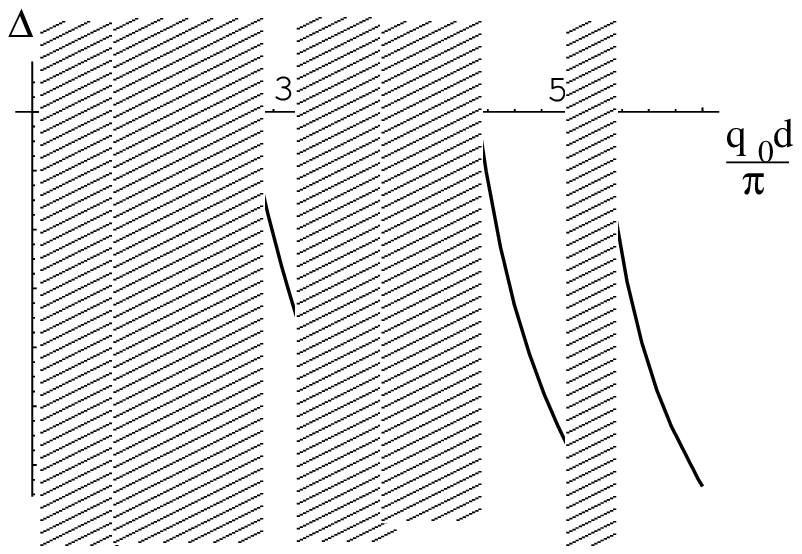}\hfill{}\includegraphics[%
  width=0.48\columnwidth]{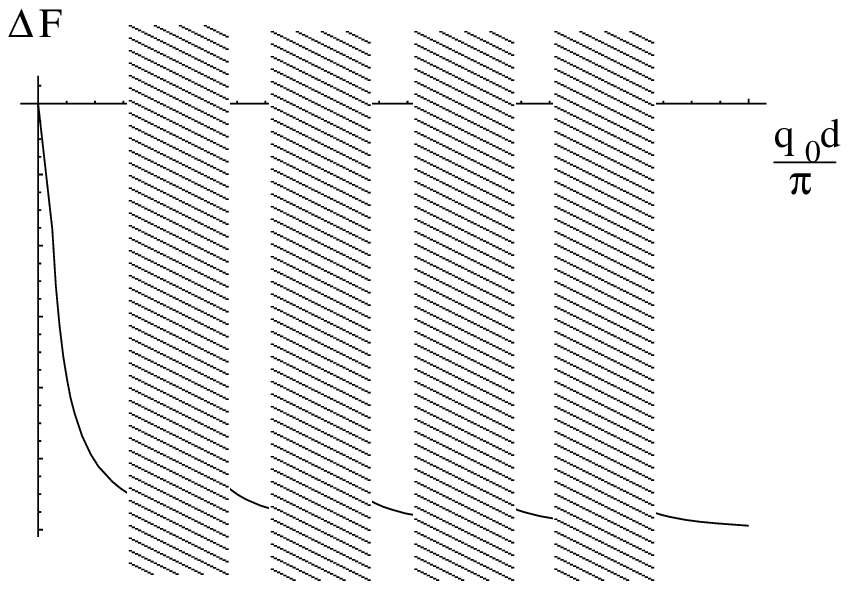}

~(a)~~~~~~~~~~~~~~~~~~~~~~~~~~~~~~~~~~~~~~~(b)

\caption{General behaviour of the free energy difference between the striped
phase ($n_{0}=1$) and the disordered phase as a function of the ratio
$q_{0}d/\pi$. The temperature in (a) is greater than the temperature
in (b). The shaded regions indicate that the ordered state is not
metastable (temperature above the spinodal) and therefore it does
not make sense to define a free energy difference.\label{fig:general_stripe}}
\end{figure}

From the definition of $\Gamma_{n_{0}}$, equation (\ref{def_gamma}),
it is clear that when the ratio $q_{0}d/\pi$ is an integer the factor
diverges. However, this does not mean that the free energy difference
diverges, since the spinodal limit also depends upon $\Gamma_{n_{0}}$.
A plot of this energy as a function of the ratio $q_{0}d/\pi$ for
a given temperature is shown in figure \ref{fig:general_stripe}.
Firstly, it is clear that there are barriers in the energy profile
whenever:

\begin{equation}
\frac{q_{0}d}{\pi}\in\mathbb{Z}\label{condition_integer}\end{equation}

\begin{figure}
\begin{center}\includegraphics[%
  width=0.70\columnwidth]{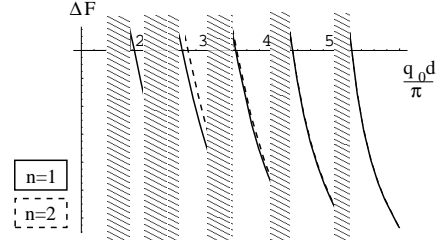}\end{center}

\caption{General behaviour of the free energy difference between the ordered
and the disordered states concerning the striped phase with $n_{0}=1$
(full line) and the striped phase with $n_{0}=2$ (dashed line). The
shaded regions indicate that the ordered state is not metastable (temperature
above the spinodal) and therefore it does not make sense to define
a free energy difference.\label{fig:compara_n0}}
\end{figure}

We also note that, as the temperature decreases, the barriers heights
become approximately uniform. Another important consequence of the
$\Gamma_{n_{0}}$ dependence upon $n_{0}$ is the break of the semi-ellipsis
degeneracy. In a mean-field approach, all the different modulations
that can exist for a certain slab's width $d$ would have the same
energy. However, as shown in figure \ref{fig:compara_n0}, when we
take into account the fluctuations, each modulation $n_{0}$ assumes
distinct energy values, implying that the high degeneracy of the minimum
energy is broken. Moreover, as the slab's width increases ($d\rightarrow\infty$),
the energies get closer again, what agrees with the result for the
infinite slab, where there is no degeneracy break. 

A deeper analysis reveals that, in fact, only the degeneracy along
the $q_{x}$ direction is broken, and not the other along the $q_{y}$
direction. This is only a reflection of the translational invariance
break along the $x$ direction, due to the existence of edges. 

It is important to analyze carefully the energy barriers that appear
when condition (\ref{condition_integer}) is met, because in such
situation (and only in it) genuine stripes, characterized by modulation
only along the $x$ direction, can appear. This configuration is given
by

\begin{eqnarray}
\left\langle m\right\rangle _{n,q} & = & A\delta_{n,n_{0}}\delta_{q,0}\\
\left\langle m\left(\vec{r}\right)\right\rangle  & = & A\sin\left(\frac{n_{0}\pi x}{d}\right)\label{stripe_qy0}\end{eqnarray}
and does not obey to the same Hartree or state equations of the configurations
previously considered. Indeed, a direct substitution of (\ref{stripe_qy0})
in (\ref{hartree_geral}) and (\ref{conjugated_field}) imply that:

\begin{eqnarray}
r & = & r_{0}+\frac{\Gamma_{n_{0}}u}{\sqrt{r}}+\frac{9}{4}uA^{2}\,,\nonumber \\
h & = & rA-\frac{3}{2}uA^{3}\,.\label{qy0_eq_aux}\end{eqnarray}

Hence, the self-consistent equation is given by:

\begin{equation}
-\frac{r_{A'}}{2}=r_{0}+\frac{\Gamma_{n_{0}}u}{\sqrt{r_{A'}}}\label{hartree_qy0}\end{equation}
and the free energy difference by:

\begin{equation}
\Delta F_{s'}=\frac{2\left(u\Gamma_{n_{0}}^{4}\right)^{1/3}}{9}\left[-\frac{\rho_{A'}^{2}}{4}-\frac{\rho^{2}}{2}-\sqrt{\rho}+\sqrt{\rho_{A'}}\right]\,.\label{energy_qy0}\end{equation}

\begin{figure}
\begin{center}\includegraphics[%
  width=0.70\columnwidth]{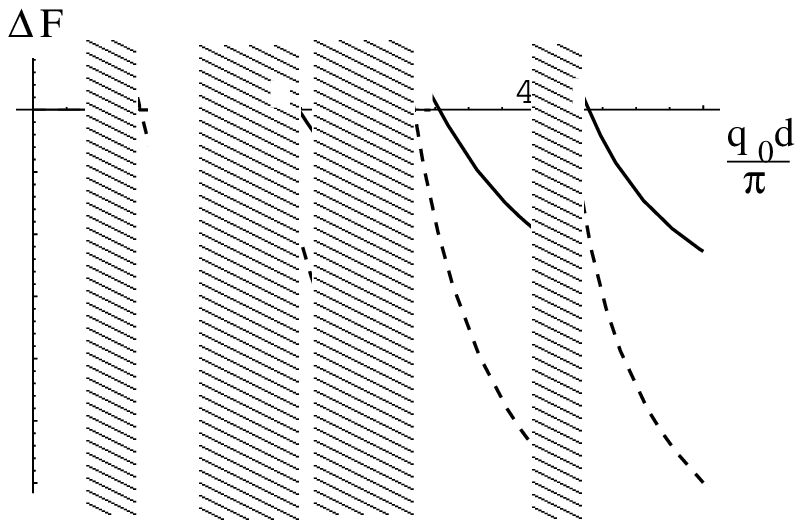}\end{center}

\caption{General behaviour of the free energy difference between the ordered
and disordered states concerning the {}``general'' striped phase
($n_{0}=1$ - full line) and the genuine striped phase with no modulation
along the $y$ direction ($n_{0}=1$ - dashed line) as a function
of the ratio $q_{0}d/\pi$. The shaded regions indicate that the ordered
state is not metastable (temperature above the spinodal) and therefore
it does not make sense to define a free energy difference.\label{fig:compara_qy=3D0}}
\end{figure}

A plot comparing the energy profiles of the {}``general'' stripes
and the stripes with no modulation along the $y$ direction is shown
in figure \ref{fig:compara_qy=3D0}. It is clear that the latter is
always more stable than the former; however, it is important to bear
in mind that the genuine stripes can only appear when condition (\ref{condition_integer})
is fulfilled. The spinodal limit for them is given approximately by
$-1.5\left(u\Gamma_{n_{0}}\right)^{2/3}$, what means that these configurations
always appear before the {}``general stripes''. Therefore, we can
say that whenever the geometric condition (\ref{condition_integer})
is met and the system can be divided in non-modulated configurations
along the $y$ direction, it will do. 

\subsection*{Bubble phases}

In an infinite slab, where the wave vector components $q_{x}$ and
$q_{y}$ are continuous, an hexagonal bubble phase is described by
the configuration \cite{doniach}:

\begin{equation}
\sum_{i=1}^{3}\cos\left(\vec{k}_{i}\cdot\vec{r}\right),\quad\sum_{i=1}^{3}\vec{k}_{i}=0\;\mathrm{and}\;\left|\vec{k}_{i}\right|=q_{0}\,.\label{bubbles_infinite}\end{equation}

It is clear that such condition can no longer be satisfied by the
finite slab, due to the boundary conditions. Than, to study other
phases than the simplest {}``striped'' ones, we consider a phase
that resembles some aspects of the bubbles in the infinite slab, as
shown in figure \ref{fig:bubble_finite}:

\begin{figure}
\includegraphics[%
  width=0.48\columnwidth]{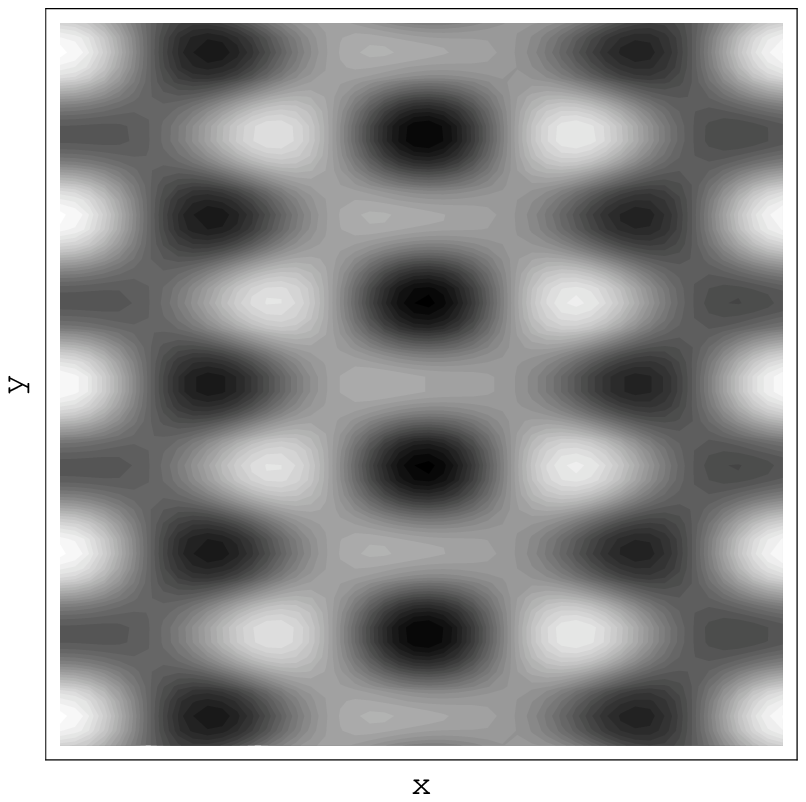}\hfill{}\includegraphics[%
  width=0.48\columnwidth]{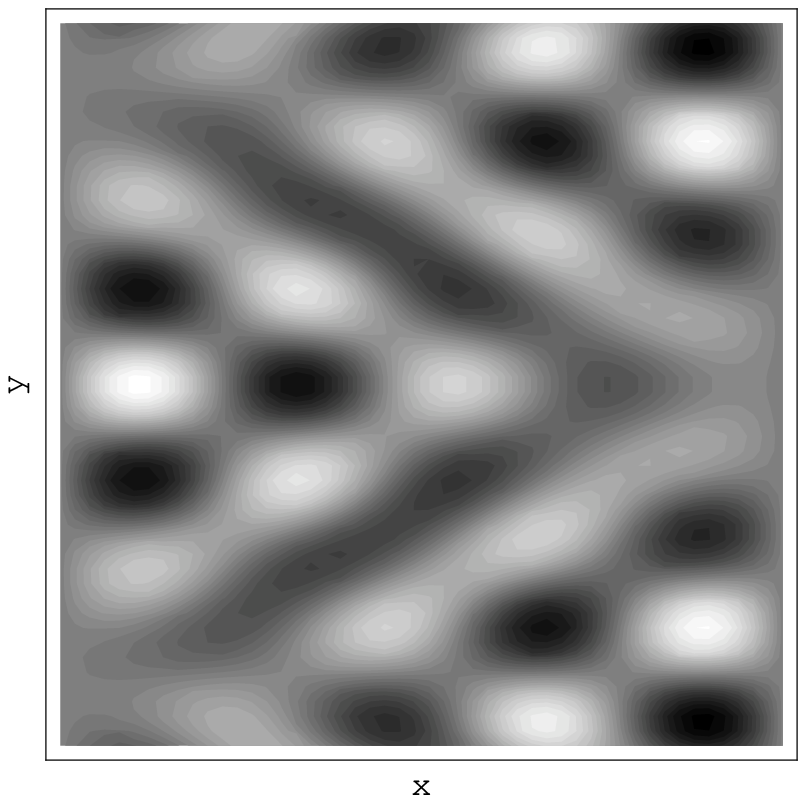}

~(a)~~~~~~~~~~~~~~~~~~~~~~~~~~~~~~~~~~~~~~~(b)

\caption{Contour plot of the order parameter along the slab's plane concerning
the {}``bubble'' phases that arise in the case of (a) an infinite
slab and (b) a finite slab.\label{fig:bubble_finite}}
\end{figure}

\begin{widetext}

\begin{eqnarray}
\left\langle m\right\rangle _{n,q} & = & A\left(\delta_{n,n_{0}}\delta_{q,q_{0y}}+\delta_{n,n_{0}}\delta_{q,-q_{0y}}+\delta_{n,n_{0}+1}\delta_{q,q_{1y}}+\delta_{n,n_{0}+1}\delta_{q,-q_{1y}}\right)\,,\nonumber \\
\left\langle m\left(\vec{r}\right)\right\rangle  & = & A\sin\left(\frac{n_{0}\pi x}{d}\right)\cos\left(q_{0y}y\right)+A\sin\left(\frac{\left(n_{0}+1\right)\pi x}{d}\right)\cos\left(q_{1y}y\right)\,,\label{bubble_finite}\end{eqnarray}

\end{widetext}where:

\begin{equation}
\sqrt{\frac{n_{0}^{2}\pi^{2}}{d^{2}}+q_{0y}^{2}}=\sqrt{\frac{(n_{0}+1)^{2}\pi^{2}}{d^{2}}+q_{1y}^{2}}=q_{0}\,.\label{bubble_modulus}\end{equation}

Once more, we adopt the name bubbles to keep the correspondence to
the case of the infinite slab, and not to describe the actual geometric
pattern. We note that this configuration can take place as long as:

\begin{equation}
\frac{q_{0}d}{\pi}\geq n_{0}+1\,.\label{condition_bubble}\end{equation}

Substituting expression (\ref{bubble_finite}) in (\ref{hartree_geral})
and (\ref{conjugated_field}) yields the following Hartree and state
equations:

\begin{eqnarray}
r & = & r_{0}+\frac{\Gamma_{n_{0}}u}{\sqrt{r}}+\frac{15}{2}uA^{2}\,,\nonumber \\
h & = & rA-\frac{9}{4}uA^{3}\,.\label{aux_eq_bubble}\end{eqnarray}

Since the conjugate field vanishes in the ordered bubble state, we
obtain, for the self-consistent equation:

\begin{equation}
-\frac{7}{3}r_{A}=r_{0}+\frac{\Gamma_{n_{0}}u}{\sqrt{r_{A}}}\,.\label{bubble_hartree}\end{equation}

The spinodal is approximately $-2.51\left(u\Gamma_{n_{0}}\right)^{2/3}$,
what means that this configuration appears after the striped phases.
In a mean-field calculation, they (and any other modulation characterized
by $q_{0}$) would appear simultaneously for both the infinite and
the finite slab. Therefore, the fluctuations break also this degeneracy,
but this is not an effect due to the finiteness of the system, since
it occurs also for the infinite slab.

Using (\ref{aux_eq_bubble}) and (\ref{bubble_hartree}), the energy
difference is calculated as:

\begin{equation}
\Delta F_{b}=\frac{4\left(u\Gamma_{n_{0}}^{4}\right)^{1/3}}{15}\left[-\frac{7\rho_{A}^{2}}{6}-\frac{\rho^{2}}{2}-\sqrt{\rho}+\sqrt{\rho_{A}}\right]\,.\label{bubble_energy}\end{equation}

\begin{figure}
\begin{center}\includegraphics[%
  width=0.70\columnwidth]{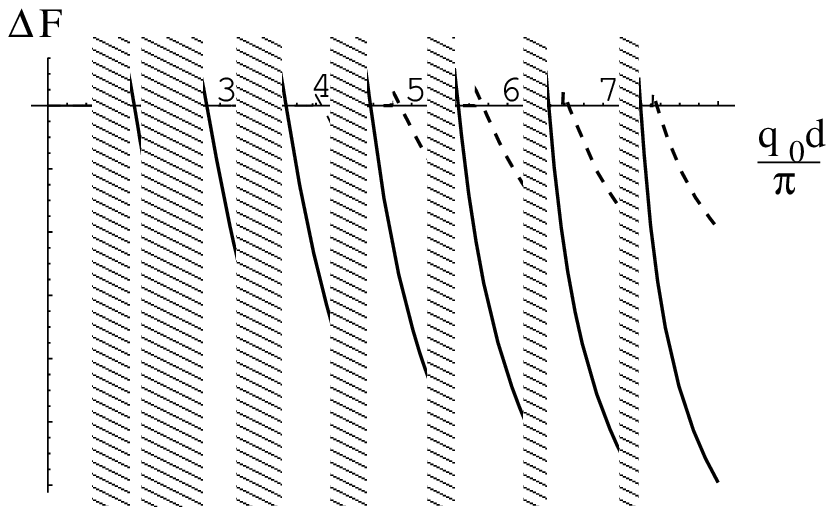}\end{center}

\caption{General behaviour of the free energy difference between the ordered
and disordered states concerning the striped (full line) and the bubble
phase (dashed line), both with $n_{0}=1$, as a function of the ratio
$q_{0}d/\pi$. The shaded regions indicate that the ordered state
is not metastable (temperature above the spinodal) and therefore it
does not make sense to define a free energy difference.\label{fig:bubbles_compare}}
\end{figure}

As it is the case for the striped phases, energy barriers are observed
when condition (\ref{condition_integer}) is met and, for different
values of the label $n_{0}$, different values of energy are observed.
Figure \ref{fig:bubbles_compare} compares the energy difference of
this bubble phase with the one referring to the striped phase; we
note that, in general, the former is greater than the latter, what
means that the simplest modulation is more stable. Moreover, since
the bubble's spinodal is lower than the stripe's spinodal, we expect
that the magnetic domain configuration will never be divided into
bubbles. 

This picture can change in the presence of an external magnetic field
along the $z$ direction. As showed by Garel and Donicah \cite{doniach}
in the case of the infinite slab, in a mean-field approach, the magnetic
field can favour the formation of bubbles instead of stripes for certain
temperature ranges. We expect that, in the present case of the finite
slab in the Hartree self-consistent approach, a similar phenomenon
can occur. However, since this is not the scope of this work, we do
not investigate further such subject .

More complex patterns built up from other combinations of the semi-ellipsis
points are also possible; however, the calculations involved become
more difficult. From the previous analysis, we expect that the simplest
modulation will be the most stable one, as it is the case for the
infinite slab. The main difference is that, in the latter case, the
spinodal of more complex phases are greater, and not lower, than the
spinodal of the simplest modulation.

\section{Applications to $MnAs:GaAs$ films}

The aspects presented in the last section are particularly interesting
on systems in which the slab's width $d$ can be varied. In fact,
this is the case for MnAs thin films grown over GaAs substrates, where
it is observed the formation of ferromagnetic terraces whose widths
depends almost linearly upon the temperature \cite{paniago}: 

\begin{equation}
d(T)=600-12T\,,\label{width_temp}\end{equation}
where $d$ is given in nanometers and $T$ in Celsius degrees. This
is valid in the region where the ferromagnetic terraces coexist with
the paramagnetic stripes, from $0\,^{\circ}\mathrm{C}$ to $50\,^{\circ}\mathrm{C}$.
In this section, we intend to discuss the domain structures inside
the ferromagnetic terraces, considering them as finite slabs, and
compare to experimental results. 

It is important to notice that, in the MnAs:GaAs system, the spins
responsible for the magnetism are not scalar (Ising-like), but vector
(due to the crystalline field, they would be better described by a
$xz$ model, and an approach following the lines of \cite{pokrovsky}
would be necessary). Besides, the film thickness is larger than $100\,\mathrm{nm}$,
what means that three dimensional domains could be formed. Nonetheless,
as we are concerned with the general picture of the problem, we believe
that this simple model proposed can outline some general properties
due to the nature of the competing interactions (strong short-range
versus weak long-range) and to the geometry involved (Dirichlet boundary
conditions in Cartesian coordinates). However, the specific features
of the domains that would be formed can be much more complex, as we
showed previously for the case of one-dimensional Néel walls \cite{fernandes}.

First of all, we need to estimate the order of magnitude of the parameters.
We do not intend to obtain an exact quantitative description, but
rather some qualitative insights about the domain structure of each
ferromagnetic terrace. Therefore, based on the experimental studies
regarding MnAs:GaAs thin films \cite{kaganer,ney,engel,paniago,iikawa},
we take $a=5\,\textrm{Å}$, $g=3$, $D=130\,\mathrm{nm}$ and $T_{c}=32\,\mathrm{meV}$. 

\begin{figure}
\begin{center}\includegraphics[%
  width=0.70\columnwidth]{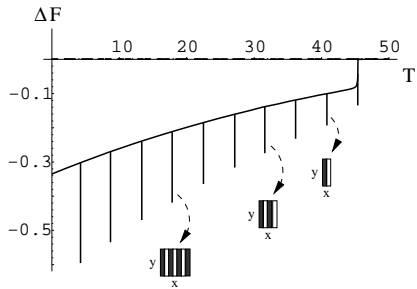}\end{center}

\caption{Free energy difference between the striped and the disordered phases
for MnAs:GaAs films as a function of temperature (in Celsius degrees).
The sharp minima refer to the configurations for which there is no
modulation along the $y$ direction and $n$ domains along the $x$
direction, where $n$ goes from $10$ (first minimum) to $1$ (last
minimum). Contour plots for some of these configurations are presented,
as well as a sketch of this profile in the case of a real system.\label{fig:mnas_stripe}}
\end{figure}

Substituting these values in the equations deduced in the previous
sections, we can study the behaviour of striped and bubble phases
inside the ferromagnetic terraces. Figure \ref{fig:mnas_stripe} shows
the energy difference between the striped phases and the paramagnetic
(disordered) phase. We note that the temperature is always below the
spinodal limit and that there are local energy minima referring to
the configurations in which there is no modulation along the $y$
direction. Such configurations comprehend structures from $1$ (the
last energy minimum) to $10$ (the first energy minimum) {}``spread''
domains lying along the $x$ direction. Note that, for a real system,
in which there are impurities and the various ferromagnetic terraces
do not have exactly the same width at a given temperature, these local
minima would not be so sharp and the free energy profile would be
continuous, as sketched in the figure.

We also see that, when $T\approx45\,^{\circ}\mathrm{C}$ ($d\approx550\,\mathrm{nm}$),
the semi-ellipsis at the Fourier space representing the minimum energy
does not comprehend any positive and integer $n$, what means that
the modulated state cannot arise anymore. Above this temperature,
it is likely that the interaction between neighbour ferromagnetic
terraces will play an important role to determine the new stable configurations.

Another aspect that is not represented explicitly in the figure is
that the break of degeneracy between states corresponding to different
number of domains $n$ is very weak, since the energy scales involved
are experimentally unnoticeable. This is due not only to the large
value of the slab's thickness $D$ but also due to the fact that the
temperature is far below the spinodal limit. Therefore, in the beginning,
when the temperature is just above $0\,^{\circ}\mathrm{C}$, all configurations
with $n\leq10$ and non-zero $q_{y}$ (such that the wave vector modulus
is $q_{0}$) are degenerate. As the temperature increases, the system
meets the first local free energy minimum, corresponding to $n=10$.
Hence, at this temperature, the terrace will be divided in $10$ domains
along the $x$ direction and no modulation along the $y$ direction.
In the sequence, all configurations with $n\leq9$ and $q_{y}\neq0$
are again degenerate; however, since the system was previously in
a $10$ domain configuration, it will be energetically favourable
to {}``destroy'' just one domain along the $x$ direction before
it meets the local minimum corresponding to $n=9$. Therefore, we
expect that each local free energy minimum corresponds to a change
in the number of domains along the $x$ direction, and that, for these
specific temperatures (where the free energy has these local {}``traps'')
, the configurations will be non-modulated along the $y$ direction. 

\begin{figure}
\begin{center}\includegraphics[%
  width=0.70\columnwidth]{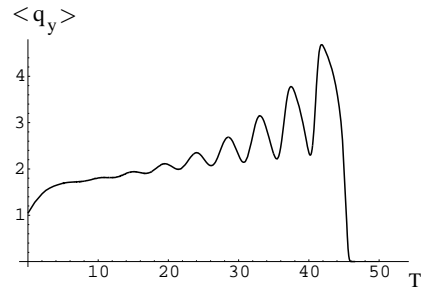}\end{center}

\caption{Mean value of the MnAs:GaAs films modulation along the $y$ direction,
in units of $10^{3}\textrm{Å}^{-1}$, as a function of the temperature
in Celisus degrees. A Gaussian distribution is considered for the
widths of the sample's terraces.\label{fig:qy_mean}}
\end{figure}

An experimental measure that can show the occurrence of these local
free energy minima is the mean value of $q_{y}$, the $y$-component
of the modulation, that can be obtained by x-ray scattering. As we
have already pointed out, in a real sample not all the ferromagnetic
slabs will have the same width determined by (\ref{width_temp}).
Instead, we can consider, as in \cite{coelho}, a Gaussian distribution
for the widths, in which the mean width is given by (\ref{width_temp}).
Figure \ref{fig:qy_mean} shows the behaviour of $\left\langle q_{y}\right\rangle $
as a function of temperature for a Gaussian distribution whose mean
standard variation is $5\%$ of the mean width. Note that the local
minima referring to the configurations whose number of domains along
the $x$ axis is large are almost suppressed, while the minima correspondent
to a small number of domains are more pronounced. 

This prediction for the qualitative behaviour of $\left\langle q_{y}\right\rangle $
is a consequence only of the competition between the interactions
involved (which generates the semi-ellipsis at the phase space) and
of the geometry considered (since the local minima appear when the
modulation length $\pi/q_{0}$ is {}``commensurate'' to the slab's
width $d$, favouring the modulation to lie only along the width's
direction). It does not depend on any particular aspect of the model,
and is expected to hold even in a vectorial model. Unfortunately,
until this moment, there are no experimental data available to verify
such prediction.

\begin{figure}
\begin{center}\includegraphics[%
  width=0.70\columnwidth]{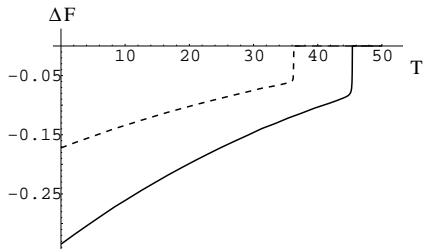}\end{center}

\caption{Free energy difference between ordered and disordered states associated
to the striped (full line) and bubble (dashed line) phases for MnAs:GaAs
films as a function of temperature (in Celsius degrees). The lines
correspondent to the configurations for which the $y$ direction is
non-modulated were removed for the sake of clearance.\label{fig:mnas_bubble}}
\end{figure}

In figure \ref{fig:mnas_bubble}, we compare the energy difference
of the bubble phases to the energy of the striped ones, but without
the local energy minima, to make the plot easier to read. It is clear
that the former is always greater than the latter, what makes one
expect to not find bubbles inside the ferromagnetic terraces. In fact,
the Magnetic Force Microscopy (MFM) images of MnAs:GaAs films do not
show configurations like bubbles, but rather structures similar to
the stripes predicted by our model and presented in figure \ref{fig:stripe}b.
Moreover, as discussed in \cite{plake,coelho}, the MFM images also
suggest that when the magnetization lies along the $z$ direction,
the ferromagnetic terraces are divided in $2$ or $3$ domains, and
not in the wider range of $1$ to $10$ domains predicted by our model.
Of course, there are several ingredients lacking in our model to make
it more realistic, like the vector nature of the magnetization and
the presence of topological defects that may take into account these
features (indeed, Garel and Doniach showed that the occurrence of
dislocations in the case of the infinite slab can melt the ordered
phase). Nonetheless, our simple model, taking into account only the
nature of the interactions and the geometry involved, is able to describe
some general qualitative features of the system.

\section{Conclusions}

In this work, motivated by the recent experiments that show the morphology
of the MnAs:GaAs ferromagnetic stripes, we developed a general theory
that describes the magnetic phase transitions at a dipolar Ising ferromagnet
slab with finite thickness and finite width. We showed that although
the modulated phase occupies a smaller volume in the $(q_{x},q_{y})$
momentum space than the one occupied by the same phase in the case
of an infinite slab, the transition between the ordered and the disordered
configurations is still first-order and induced by fluctuations (a
Brazovskii type transition). This happens because all the momentum
space {}``shrinks'' in the case of the finite slab, due to the discretization
of the momentum component in the limited direction ($x$), what is
a consequence of the boundary conditions. And what is fundamental
to the Brazovskii transition occur is not the absolute volume of the
modulated phase, but how large it is when compared to the rest of
the momentum space. So, in the case we studied there is still enough
phase space to the fluctuations of the order parameter induce a transition.

We also showed that, for the finite slab, there is the possibility
of first-order transitions between the modulated and the disordered
phases driven not by temperature, but by variation of the slab's width.
Studying two different modulated configurations, namely, striped (figure
\ref{fig:stripe}b) and bubble (figure \ref{fig:bubbles_compare}b)
ones, we showed that the first has lower energy and higher spinodal
then the latter, what means that bubble phases are not expected to
be observed in such materials. This is in qualitative agreement with
MFM images realized on MnAs:GaAs thin films, where the domain structures
inside the ferromagnetic terraces are similar to the stripes of figure
\ref{fig:stripe}b).

Another effect of the finiteness of the slab was that modulated phases
characterized by different number of {}``spread'' domains (referring
to the semi-ellipsis minimum energy projection along the $q_{x}$
axis ) have different energies. Thus, fluctuations of the order parameter,
together with finite size effects, break the high degree of degeneracy
of the ground state. In the case of the infinite slab, there was already
a break of degeneracy (referring to the differences in energy between
striped and bubble phases, for instance), but in the finite case,
we noticed that it was deeper (as some degeneracies between different
stripes are also broken). It is clear that this effect, in the momentum
space, is a consequence of the break of translational invariance in
the real space. However, for the case of the MnAs:GaAs system studied
so far, we saw that this break is practicable undetectable, due to
the large film thickness (hundreds of lattice parameters) and temperature
(that is far below the spinodal limit).

In addition, we noted that for slabs whose widths $d$ are {}``commensurate''
with the modulation wavelength $\pi/q_{0}$, the most stable configuration
is the one for which there is no modulation along the unlimited ($y$)
direction. This is reflected by the occurrence of steps in the energy
profiles of the system. In what concerns MnAs:GaAs films, we noticed
that such steps generate local minima in the free energy profile that
are responsible for changes of the number of {}``spread'' domains
in the $x$ direction inside the ferromagnetic terraces. Although
MFM images reveal that there can be phases, inside the terraces, with
different number of domains ($2$ and $3$), they show that this number
is not so large as the predicted by our model (below $10$). Moreover,
it was not reported yet any configuration without modulation along
the $y$ direction. In fact, we do not expect that this actually happens,
not only because of the precision required (the slab's width must
be commensurate to the modulation wavelength), but also for the fact
that the $y$ direction is not unlimited. Rather, we expect that the
mean modulation along the $y$ direction increases with temperature
and oscillates near the region where the {}``commensurate widths''
occur, as showed in figure \ref{fig:qy_mean}. Unfortunately, as of
yet there is no experiments that can verify this behaviour. 

Hence, this model is only the first step towards a more complete understanding
of the complex features of the MnAs:GaAs phase diagram and could be
applied to other similar systems where there is a competition between
organizing and frustrating interaction plus a finite slab geometry.
An improved theory should surely comprehend not an Ising model, but
a vectorial one, even because it is observed, in the MFM images, configurations
in which the magnetization lies along the $x$ axis. Due to the crystalline
field of the material, no components of the magnetization along the
$y$ direction is expected. Another important thing to take into account
is the topological defects, which can play fundamental role in two-dimensional
systems.

\section*{Acknowledgments}

The authors kindly acknowledge fruitful discussions with R. Magalhães-Paniago
and L. Coelho and also the financial support from CNPq and FAPESP.

\section*{Appendix A: Dipolar energy}

In this appendix, we explicitly calculate the magnetostatic energy
of an arbitrary configuration of the slab. First, let us derive a
general expression to compute magnetostatic energies: given a certain
magnetization $\vec{M}(\vec{r})$, in the absence of free currents,
the magnetic field generated can be described by the magnetic scalar
potential $\phi(\vec{r})$ that satisfies the Poisson equation \cite{jackson}:

\[
\nabla^{2}\phi=-4\pi\rho\,,\]
where $\rho$, the effective magnetic poles density, is given by:

\[
\rho=-\vec{\nabla}\cdot\vec{M}\,.\]

Hence, the magnetostatic energy is written, in the Fourier space,
as:

\begin{equation}
E=2\pi\int\frac{\left|\rho(\vec{k})\right|^{2}}{k^{2}}d^{3}k\,.\label{ap:energia_definicao}\end{equation}

Let us now apply this formalism to our specific case, namely, an arbitrary
magnetization of a slab with thickness $D$ and length $d$. Making
use of step functions, it can be written, in the whole space, as:

\begin{equation}
\vec{M}(\vec{x})=M(x,y)\theta(x)\theta(d-x)\theta(D/2-z)\theta(D/2+z)\hat{z}\,,\label{ap:magnet_todo}\end{equation}
where $M(x,y)$ is given by (\ref{magnet_fourier}), as explained
before. Hence, a straightforward calculation yields, for the effective
magnetic poles density in the Fourier space:

\begin{widetext}

\[
\rho(\vec{k})=\left(\frac{g\mu_{B}}{a^{3}}\right)\frac{-iL_{y}}{(2\pi)^{3/2}}\sum_{n}m_{n,k_{y}}\sin\left(\frac{k_{z}D}{2}\right)\left[e^{ik_{x}d}(-1)^{n}-1\right]\left[\frac{1}{(k_{x}+n\pi/d)}-\frac{1}{(k_{x}-n\pi/d)}\right]\,.\]

Substituting this expression in (\ref{ap:energia_definicao}), we
obtain the dipolar energy:

\[
E=\frac{2L_{y}^{2}}{d^{2}}\left(\frac{g\mu_{B}}{a^{3}}\right)^{2}\sum_{n,n'}nn'\int d^{3}k\,\frac{m_{n,k_{y}}m_{n',-k_{y}}}{\left(k_{x}^{2}+k_{y}^{2}+k_{z}^{2}\right)}\sin^{2}\left(\frac{k_{z}D}{2}\right)\frac{\left[1+(-1)^{nn'+1}\cos(k_{x}d)\right]}{\left(k_{x}^{2}-\frac{n^{2}\pi^{2}}{d^{2}}\right)\left(k_{x}^{2}-\frac{n'^{2}\pi^{2}}{d^{2}}\right)}\,,\]
 where the summation is to be understood as involving $n$ and $n'$
with same parity. Therefore, there are three integrals to be evaluated;
the one referring to $k_{y}$ can be rewritten as a summation and
the one referring to $k_{z}$ can be calculated analytically:

\[
\int_{-\infty}^{\infty}dk_{z}\frac{\sin^{2}\left(\frac{k_{z}D}{2}\right)}{\left(k_{x}^{2}+k_{y}^{2}+k_{z}^{2}\right)}=\frac{\pi}{2}\frac{\left(1-e^{-D\sqrt{k_{x}^{2}+k_{y}^{2}}}\right)}{\sqrt{k_{x}^{2}+k_{y}^{2}}}\,,\]
yielding the following expression for the density of dipolar energy:

\[
f_{dip}=\frac{E}{L_{y}Dd}=\frac{4\pi^{2}}{Dd^{3}}\left(\frac{g\mu_{B}}{a^{3}}\right)^{2}\sum_{k_{y},n,n'}nn'm_{n,k_{y}}m_{n',-k_{y}}\int_{0}^{\infty}dk_{x}\frac{\left(1-e^{-D\sqrt{k_{x}^{2}+k_{y}^{2}}}\right)\left[1+(-1)^{nn'+1}\cos(k_{x}d)\right]}{\sqrt{k_{x}^{2}+k_{y}^{2}}\left(k_{x}^{2}-\frac{n^{2}\pi^{2}}{d^{2}}\right)\left(k_{x}^{2}-\frac{n'^{2}\pi^{2}}{d^{2}}\right)}\]
which, by a simple change of coordinates, becomes expression (\ref{desmagnet_energ}).\end{widetext}


\begin{thebibliography}{10}
\bibitem{science}M. Seul and D. Andelman, Science \textbf{267}, 476 (1995)
\bibitem{doniach}T. Garel and S. Doniach, Phys. Rev. B \textbf{26}, 325 (1982)
\bibitem{allen}R. Allenspach and A. Bischof, Phys. Rev. Lett. \textbf{69}, 3385 (1992)
\bibitem{wu}D. Wu, D. Chandler and B. Smith, J. Phys. Chem. \textbf{96}, 4077
(1992)
\bibitem{langmuir}R. M. Weis and H. M. McConnell, Nature (London) \textbf{310}, 47 (1984)
\bibitem{fredrickson}G. H. Fredrickson and E. Helfand, J. Chem. Phys. \textbf{87} (1987)
\bibitem{finito_pbc1}E. Brézin and J. Zinn-Justin, Nucl. Phys. B \textbf{257}{[}FS14{]},
867 (1985)
\bibitem{finito_pbc2}S. Singh and R. K. Pathria, Phys. Pev. B \textbf{34}, 2045 (1986)
\bibitem{finito_pbc3}A. Esser, V. Dohm and X. S. Chen, Physica A \textbf{222}, 355 (1995)
\bibitem{finito_Dbc1}J. Rudnick, G. Gaspari and V. Privman, Phys. Pev. B \textbf{32}, 7594
(1985)
\bibitem{finito_Dbc2}W. Huhn and V. Dohm, Phys. Rev. Lett. \textbf{61}, 1368 (1988)
\bibitem{finito_Dbc3}V. Dohm, Z. Phys. B - Condensed Matter \textbf{75}, 109 (1989)
\bibitem{braso}S. A. Brazovskii, Zh. Eksp. Teor. Fiz. \textbf{68}, 175 (1975)
\bibitem{mnas_prl}A. K. Das, C. Pampuch, A. Ney, T. Hesjedal, L. Däweritz, R. Koch and
K. H. Ploog, Phys. Rev. Lett. \textbf{91}, 087203 (2003)
\bibitem{tanaka}M. Tanaka, Semicond. Sci. Technol. \textbf{17}, 327 (2002)
\bibitem{bean}C. P. Bean and D. S. Rodbell, Phys. Rev. \textbf{126}, 104 (1962)
\bibitem{kaganer}V. M. Kaganer, B. Jenichen, F. Schippan, W. Braun, L. Daweritz and
K. H. Ploog, Phys. Rev. B \textbf{66}, 045305 (2002)
\bibitem{ney}A. Ney, T. Hesjedal, C. Pampuch, A. K. Das, L. Daweritz, R. Koch,
K. H. Ploog, T. Tolinski, J. Lindner, K. Lenz and K. Baberschke, Phys.
Rev. B \textbf{69}, 081306 (2004)
\bibitem{engel}R. Engel-Herbert, J. Mohanty, A. Ney, T. Hesjedal, L. Däweritz and
K. H. Ploog, Appl. Phys. Lett. \textbf{84}, 1132, 2004.
\bibitem{paniago}R. Magalhães-Paniago, L. N. Coelho, B. R. A. Neves, H. Westfahl, F.
Iikawa, L. Daweritz, C. Spezzani and M. Sacchi, Appl. Phys. Lett.
\textbf{86}, 053112 (2005)
\bibitem{iikawa}F. Iikawa, M. Knobel, P. V. Santos, C. Adriano, O. D. D. Couto, M.
J. S. P. Brasil, C. Giles, R. Magalhães-Paniago and L. Daweritz, Phys.
Rev. B \textbf{71}, 045319 (2005)
\bibitem{negele}J. W. Negele and H. Orland, \emph{Quantum Many-Particle Systems},
Advanced Book Classics, Perseus, USA (1998)
\bibitem{lubenski}P. M. Chaikin and T. C. Lubensky, \emph{Principles of Condensed Matter
Physics}, Cambridge University Press, United Kingdom (1995)
\bibitem{pokrovsky}A. Kashuba and V. L. Pokrovsky, Phys. Rev. Lett. \textbf{70}, 3155
(1993)
\bibitem{fernandes}R. M. Fernandes, H. Westfahl Jr., R. Magalhães-Paniago and L. N. Coelho,
cond-mat/0605357 (2006)
\bibitem{plake}T. Plake, T. Hesjedal, J. Mohanty, M. Kastner, L. Däweritz and K.
H. Ploog, Appl. Phys. Lett. \textbf{84}, 1132 (2004)
\bibitem{coelho}L. N. Coelho, R. Magalhães-Paniago, B. R. A. Neves, F. C. Vicentin,
H. Westfahl, R. M. Fernandes, F. Iikawa, L. Däweritz, C. Spezzani
and M. Sacchi, cond-mat/0605356 (2006)
\bibitem{jackson}J. D. Jackson, \emph{Classical Electrodynamics}, John Wiley and Sons,
USA (1998)\end{thebibliography}
\end{document}